# Determining Absence of Unreasonable Risk:
# Approval Guidelines for an Automated Driving System Deployment

Favarò F.M., Schnelle, S., Fraade-Blanar, L., Victor, T., Peña, M., Webb, N.,
Broce, H., Paterson, C., Smith, D. - *Waymo LLC*

**Abstract:** This paper provides an overview of how the determination of absence of unreasonable risk can be operationalized. It complements previous theoretical work published by existing developers of Automated Driving Systems (ADS) on the overall engineering practices and methodologies for readiness determination. Readiness determination is, at its core, a risk assessment process. It is aimed at evaluating the residual risk associated with a new deployment. The paper proposes methodological criteria to ground the readiness review process for an ADS release. While informed by Waymo's experience in this domain, the criteria presented are agnostic of any specific ADS technological solution and/or architectural choice, to support broad implementation by others in the industry. The paper continues with a discussion on governance and decision-making toward approval of a new release candidate for the ADS. The implementation of the presented criteria requires the existence of appropriate safety management practices in addition to many other cultural, procedural, and operational considerations. As such, the paper is concluded by a statement of limitations for those wishing to replicate part or all of its content.

**Keywords:** Automated Driving Systems, Safety Determination, Acceptance Criteria, Decision-Making, Safety Assurance, Risk Management.

---

## Introduction

The feasibility and acceptance of Automated Driving Systems (ADS) [1] technology hinges on the ability to rigorously and comprehensibly detail how the determination of safety has been carried out. A number of publications and books on this topic (see, for example [2], [3], [4], [5], [6]) have informed the discussion across industry, academia, and policy-makers, and shaped the latest thinking that is being gradually codified into recommended practices [7], technical specifications [8], standards [9], and regulatory publications [10] [11].

In the context of Level 4 ADS dedicated vehicles, which operate without the presence of a fallback ready user [1] (referred to, in this paper, as "Rider Only" - RO - operations, but also dubbed "Driver-Out" or "no-user-in-charge"), the current debate has often centered on the false choice [12] between: (1) aggregate measures of safety outcomes (e.g., quantitative statistical estimates of effectiveness to reduce injury risk [13] [14] [15] and associated limitations); and (2) principled measures able to provide adequate evidence of acceptable behavior in single events/scenarios (e.g., ability to comply with rules of the road). Waymo, at this time the only company operating at-scale ADS in rider-only (RO) operations within the United States, has long advocated for a combination of both aggregate and event level types of indicators [16] [17]: both are necessary to determine that the level of residual risk posed by the system remains acceptable, and their presentation as a dichotomy or contrasting approaches has hindered the understanding of the broader ADS community. We further argue that contraposing these two types of measures (i.e., at the aggregate and at the event level) in light of the long-standing philosophical debate across utilitarian and deontological perspectives[1] is a mischaracterization of stakeholders' positions that is naturally resolved through a more holistic approach to safety [17].[2] Put more simply, the combination of both perspectives is essential to a holistic approach to safety, which thus needs to be based on multiple complementary methods, metrics, and supporting processes.

A factor much less discussed, but equally important, is the complementarity between prospective and retrospective safety analyses, where the prospective, *pre-deployment* qualification of a Level 4 ADS (based on estimates obtained from simulation, closed-course, and public road testing) is eventually confirmed by retrospective measures (obtained from *post-deployment observations*) for a specific operational design domain. Those activities are unfortunately often confused or not sufficiently disambiguated, leading to further mischaracterization of approval processes currently adopted within the industry.

---

[1] Simply put, utilitarianism is a philosophical theory that argues, when presented with a choice, for choosing the option that maximizes benefits/outcomes for a population of interest. This is often simplified into tenets like "the greatest good for the greatest number of people". Conversely, deontology represents the philosophical practice of evaluating the moral and ethical implications of the actions associated with each choice being evaluated in a principled-driven rationalization of options.

[2] We do not intend here to dismiss the complexity of the ethical and moral questions connected to the deployment of ADS technology and the determination of appropriate on-road behavior. The point is rather to remark that an approach that *combines* rather than *contraposes* different types of indicators is better suited for the evaluation of safety [16].

In 2023, Waymo presented [16] its approach to the creation of a safety case, detailing the importance of tackling the question of determining absence of unreasonable risk (the current definition of safety according to industry [18] and regulatory [19] standards) in a manner that sufficiently demonstrates competence to operate on public roads. That presentation was grounded in standardized definitions of acceptance criteria and validation targets [20], detailing a theoretical framing to encompass a layered hazard-based risk assessment within the dynamic and iterative development of an ADS release process. Other ADS companies also shared information about their approach to safety, including details on the structure and topics of their safety case (see Aurora's Safety Case Framework), and/or commitment for validation of the safety determination (see Gatik's Safety Case Assessment Framework).

Across the preceding literature, the question of how to *operationalize* the determination of Absence of Unreasonable Risk (AUR) for an ADS remained, however, unspecified. Thus the desire, in this paper, to provide a proposal for a set of criteria for readiness determination and approval process of an ADS.

According to state of the art industry practices [18], the evaluation of whether residual risk meets acceptable thresholds can be based on a number of criteria, be they connected to performance or processes and their associated rigor. This idea is at the basis of the determination of AUR [18]. The set of acceptance criteria proposed here is mapped to the core methodologies that shape ADS readiness determination at Waymo [21]. While their internal use at Waymo has remained reasonably stable (and their definition is done at a level of abstraction that enabled such stability) it is understood that the set of criteria may not be set in stone, and that any substantial methodological changes and/or the addition of novel evaluation approaches may alter the set of criteria being considered as well as the associated metrics. As such, this publication is intended as a proposal for AUR acceptance criteria that still enables justification of departures and/or equivalency claims through different measures. Based on our experience, revisions of the set of criteria are tied to the scale and maturity of a company, as the availability and reliance on data from previously fielded ADS releases may calibrate and fine-tune a company's approach to approval. Future revisions may also be possible as informed by the continuous evolution of an ADS (e.g., the addition of a new use-case).

As detailed in [16] and [21], Waymo's readiness approval is anchored in its Safety Framework: a set of principles, methodologies, and governance structure on which the qualification of each release candidate is based. Our Safety Case, conversely, fulfills safety assurance functions by first documenting and explaining, and then querying and pressure testing the validity and implementation of said Safety Framework. Our Safety Case, in other words, provides the evidentiary foundation to support the conclusion that consistent and rigorous application of the principles, methodologies, and governance structure will accurately ensure that the subject ADS will meet the overall criterion of AUR both at its introduction and through subsequent iterations embodied in software releases. The methodological criteria presented in this paper were developed in accordance with Waymo's Safety Framework (as first introduced to the public in 2020 [21]), and reflect what is documented in Waymo's Safety Case [16]. Yet, they have been abstracted and generalized toward a possible industry-wide use. Furthermore, a discussion of governance is undertaken, detailing how the provided information is leveraged for decision-making toward approval of a deployment.

Targets and quantitative benchmarks are not included in this paper. Minimum benchmarks will need to be agreed upon broadly by industry and regulators, but only after a more precise conversation on which metrics and indicators can be used to aggregate residual risk is undertaken. As such, this paper offers a first step toward standardization (and, as applicable, toward informing future regulation) on the determination of absence of unreasonable risk. If sought, information pertaining to targets' setting for some of the current methodologies employed at Waymo can be found in other publications (see examples in [22], [23], [24] and [30]).

Finally, this paper does not cover, as explained in more detail below, design activities and processes that ground the rigorous development of an ADS ahead of the actual approval process. It is focused instead on the product evaluation and readiness determination that occurs at discrete points in time to approve the deployment of a candidate ADS configuration[3].

## Readiness Review Criteria

### *High-Level Structure*

Table 1 presents an overview of twelve methodological criteria: each criterion is leveraged during Waymo's readiness review for a given ADS candidate configuration. The table is organized according to the section headings presented in [21], as shown in the first column. This first column is anchored on a number of widely-understood activities (e.g., scenario based testing, public road testing) that collectively shape ADS safety determination in well-known standards and industry-best practices [7, 8, 18, 20]. As can be gleaned from the structure of Table 1, the mapping between each criterion and the headings listed in the 2020 paper is not one-to-one: certain criteria span multiple types of testing activities (e.g., the analysis of predicted collisions spans both on-road testing, monitoring, and evaluation as well as simulated deployments) and/or, vice-versa, certain types of approaches (e.g., scenario-based verification in the first column) correspond to multiple methodologies and criteria (e.g., those to test collision avoidance capabilities; a portion of those to test traffic rules compliance proficiency; a portion of the analyses dedicated to Vulnerable Road Users interactions; etc.).

Table 1. Overview of the criteria included in this paper, mapped to the activities first presented in [21].

| Mapping to Activities presented in [21] | Methodological Criteria toward the Determination of Absence of Unreasonable Risk | | | | |
|---|---|---|---|---|---|
| **Hazard Analysis** | (1) System Safety Criterion | | | | |
| **Cybersecurity** | (2) Cybersecurity Criterion | | | | |
| **Verification and Validation** Inclusive of: Base Vehicle, Motion Control, Sensing, Computational platform, Fault Detection & Response | (3) Verification and Validation Criterion | | | | |
| **Scenario-based Verification** | (4) Collision Avoidance Testing Criterion | | (7) Rules of the Road Compliance Criterion | (8) Vulnerable Road Users Interactions Criterion | (9) High Severity Assessment Criterion |
| **Simulated Deployments** | (5) Predicted Collision Risk Criterion | (6) Impeded Progress Criterion | | | |
| **On-road Testing & Monitoring*** | | | | | |
| **Risk Management** | (10) Conservative Severity Estimates Criterion | | | | |
| | (11) Risk Management Criterion | | | | |
| **Field Safety & Fleet Ops** | (12) Field Safety Criterion | | | | |

***This activity was separated out from the general category of "fleet operations".*

Within the following section, for each criterion we present a natural language description, detailing both the safety objective and the evaluation approach undertaken. The goal of this statement is to specify how the criterion contributes to the overall determination of absence of unreasonable risk.

As noted in [16], within Waymo's approach each acceptance criterion may contribute to the analysis of safety according to one or more layers across the architectural, behavioral, and operational domains.[4]  In the following

---

[3] This entails a specified configuration of both the hardware and the software across all layers (architectural, behavioral, and operational) as integrated into a specific vehicle platform for a specific use-case (e.g., ride-hailing).

[4] In [16], the authors presented the detailed decomposition for the behavioral layer, using a visual pie chart. The architectural and operational layers feature the same categories of "severity potential", "ADS functionality", and "level of aggregation" presented for the behavioral layer in [16]. They also consider specific domains of analysis for those layers. For the

section, we call out for each criterion the particular layers it contributes to at a high level. A more precise mapping had been exemplified in [16] for the behavioral layer only. Such mapping highlights the contribution from a particular methodology to a *predefined evaluation space* considered for each of the three layers [16] and enables the *holistic assessment of the collection of criteria* (analyzed as a set, rather than on an individual basis). As noted in all Waymo publications since [21], no methodology, on its own, is capable of providing appropriate coverage with adequate confidence of the entirety of the evaluation space; rather, when signal from multiple methodologies are overlaid, a more comprehensive picture of residual risk is obtained, where "signal" is here intended to combine both quantitative and qualitative inference measures yielded by the multiple indicators employed to evaluate each methodology criterion. This approach, termed *"signal-based evaluation"*, in turn enables a more systematic *analysis of residual risk for a complex system*, where a fully quantitative aggregate roll-up of risks coming from multiple/diverse potential sources of harm is not currently achievable.

## Criteria Presentation

### (1) System Safety:

*Risk Acceptance Criterion*: A safety analysis and risk assessment process is conducted in relation to architectural, behavioral, and in-service operational hazards applicable to the ADS. Subsystem and component level hazard causes stemming from malfunctions, functional insufficiencies, insufficiency of specification, and reasonably foreseeable misuse are identified in safety analyses to attain an estimate of inherent system risk, based on probability of occurrence and worst-plausible severity of harm for each unmitigated cause. Risk targets are specified to identify and prioritize mitigations for each identified hazard cause that exceeds the established risk target. Completion (e.g., tracking percentage coverage) is determined upon execution of all in-scope safety analyses, and implementation and verification of safety requirements resulting from the analyses. Additionally, an assessment of maturity evaluates the breadth and depth of coverage of the safety analyses, confidence in the verification process, and overall health and rigor of the methodology. The given context[5], along with the types of hazards and applicable triggering mechanisms, influence the type of safety analyses chosen. Maps to: architectural, behavioral, operational layers.

### (2) Cybersecurity:

*Risk Acceptance Criterion*: Acceptance is predicated on assessing and mitigating risks identified during either ongoing threat and vulnerability identification or active events. An assessment of risk from identified vulnerabilities for the ADS is carried out considering: (a) the possibility that an attack will locate an inroad and succeed in exploiting vulnerabilities in the Service, based on the difficulty of developing and executing such an attack; and (b) the severity of the potential consequences and impact of such an attack on the current operation, other road users or members of the public, and/or the ADS operating entities. Additionally, the probability of an attacker attempting such an effort, given the potential threat actors motivated to attack the Service may be evaluated based on availability of such information. For vulnerabilities assessed as posing substantial Cybersecurity risk (e.g., through a multi-dimensional risk matrix evaluating possibility vs. impact), a vulnerability management process of detailed analysis and mitigation and/or referral and/or escalation to other stakeholders is followed. Maps to: architectural layer.

### (3) Verification & Validation:

*Risk Acceptance Criterion*: Potential sources of harm are managed through the definition and verification of requirements associated with architectural, behavioral, and in-service operational functions. Requirements are generated by subject matter experts utilizing well documented processes and tools (e.g., according to

---

architectural layer those include: the base vehicle, the sensing hardware and firmware, the compute and overall reasoning architectural choices, and related interfaces; for the in-service operational layer those include: rider interaction, fleet response (inclusive of remote assistance, roadside assistance, and other mission management functions), and maintenance and depot management.

[5] Within Waymo's approach, the context is defined as the vehicle and operational configuration (see footnote 3), the ODD in scope for the release, the targeted locations of deployment, and the deployment size [16].

predefined categories and level of refinement to inform requirements' coverage). Completion of the verification activities is based on passing measurable and explicit verification items that are attached to each requirement. Residual risk is informed by: 1) analyzing the safety risk of all requirements without passing verification, where risks that are assessed to be above "Very Low" risk are tracked through the Risk Management[6] processes, 2) feedback and in use monitoring from other methodologies that speak to either the need to update requirements or the need for new requirements. The given context influences the type and number of requirements in scope for the release. Maps to: architectural, behavioral, operational layers.

### (4) Collision Avoidance Testing:

*Risk Acceptance Criterion*: The predicted collision avoidance capability attained by the ADS in a number of conflict scenarios initiated by the actions of other road users is assessed through a comparison with a chosen behavioral reference benchmark.[7] Scenario groups are graded at an aggregate level, with individual scenarios within a group contributing to a neutral/positive/negative gap for the ADS when it shows even/better/worse performance than the behavioral benchmark in terms of collision outcomes (e.g., binary output) and severity (e.g., injury likelihood). Minimum passing scores vary between scenario-specific groups, each including either vehicle-to-vehicle or vehicle-to-vulnerable-road-users interactions. Maps to: behavioral layer.

### (5) Predicted Collision Risk:

*Risk Acceptance Criterion*: The predicted ADS performance in terms of collision rates per million miles, stratified by severity (e.g., breakdown in levels as in [18]), is assessed through a comparison to an appropriately adjusted[8] human-derived benchmark. The aggregate-level predicted collision rate combines observed and counterfactually simulated collisions and near collision events collected during testing and past operations of the ADS. For the given context, including the sought scale of deployment, an assessment of residual risk is based on: (a) confidence in the computation of the delta resulting from the comparison with the human-derived benchmark; (b) evidence of continuous improvement based on analysis of regression of predicted collision performance of the present ADS release compared to prior ones and mapped to tracked engineering developments; and (c) an assessment of the ADS behavior for specific interaction types. Maps to: behavioral layer.

### (6) Impeded Progress:

*Risk Acceptance Criterion*: The ADS predicted performance for improper stops, strandings[9] and slowings is assessed through a comparison to specified target rates (e.g., duration normalized by exposure) for: (a) situations near routes that allow high-speed travel; (b) delays to other vehicles; (c) situations in which a mission cannot be completed or the ADS is unable to accept new missions and is recovered from a location outside an ADS depot; and (d) (bounded) estimated contribution to predicted collision rates. The given context, including the sought scale of deployment, affects how the benchmark rates are set and how they influence the determination of safety. Maps to: behavioral layer.

### (7) Rules of the Road Compliance:

*Risk Acceptance Criterion*: The ADS predicted performance with respect to rules of the road compliance is assessed through: (a) thorough identification and interpretation of applicable road rules so that the rules can be translated into testable functional requirements for an automated driving system; (b) for such internal requirements, achievement of minimum passing scores across pre-specified scenario groups that establish proficiency through foundational coverage; (c) a comparison between specified targets and the ADS predicted performance (e.g., rates of events per mile or success rates per encounter) for prioritized compliance behaviors, estimated from field monitoring of on-road operations or equivalent simulated deployments. The risk assessment is complemented by a qualitative analysis of the events for a selected sample of interest. Maps to: behavioral layer.

---

[6] See criterion (11).

[7] For example, the NIEON (non-impaired, eyes on conflict) model provides a benchmark used internally at Waymo [23, 24].

[8] The intent here is to ensure applicability of the benchmark for an apples-to-apples comparison. See [25].

[9] A stop is here intended as a temporary recoverable situation, while a stranding would lead to vehicle retrieval.

## (8) Vulnerable Road Users Interactions:

*Risk Acceptance Criterion*: The predicted aggregate performance of the ADS in terms of behavioral capabilities displayed in close proximity to Vulnerable Road Users* (VRU) is assessed through a qualitative analysis that establishes commensurate performance with context-dependent targets. The analysis evaluates the cross-section of notable events involving VRU interactions, sourced from the evaluation of all acceptance criteria, and complemented by customized analyses of events' clusters and interaction exposure. Measured performance against each individual criterion target (e.g., vehicle-to-VRU collision rates or collision avoidance gap scores with respect to a behavioral benchmark) is complemented by a summary of aggregated ADS performance. Maps to: behavioral layer. **Pedestrians, cyclists, turboped (i.e., scooterists) and motorcyclists are considered within the VRU category.*

## (9) High Severity Assessment:

*Risk Acceptance Criterion*: An assessment of high-severity risks stemming from the ADS behavior is informed by: (a) a coverage assessment of reasonably foreseeable high injury potential scenarios (HIPS); (b) a qualitative confidence assessment of the ability to minimize the unknown HIPS space (e.g., rate of discovery of new HIPS); (c) a performance assessment of the ADS predicted performance in HIPS (e.g., collision outcomes, contributory role). The High Severity assessment is organized by a conflict typology and provides actionable feedback to identify potential gaps in existing coverage and improve the ADS performance in high injury potential scenarios prioritized by frequency (>X% per grouping). Depending on context, requirements on behaviors in key high severity situations are evaluated to ensure performance demonstrates sufficient driving proficiency for both initiator and responder roles and meets minimum coverage specifications. Maps to: behavioral layer.

## (10) Conservative Severity Estimates:

*Risk Acceptance Criterion*: A risk assessment process is carried out to investigate the root-cause and oversee the disposition of bugs escalated from a range of sources including on-road driving and virtual testing using simulation tools. Risk stemming from each bug is evaluated and bugs are routed by analyzing safety context-specific worst-plausible (conservative) severity, root causes, impact to the rest of the system, and novelty. Completion (e.g., tracking percentage coverage) is determined when all in-scope issues/bugs have been resolved through appropriate disposition. Maps to: architectural, behavioral layers.

## (11) Risk Management:

*Risk Acceptance Criterion*: Sources of risk arising from either missing or insufficiently covered requirements are evaluated based on a gap analysis between intended vs. expected (from observation or simulation) ADS behaviors. Risk is then assessed according to a context-specific matrix featuring likelihood and severity of harm resulting from potential collisions and other accidents. Individual risks considered to remain above the "very low" level following the implementation of mitigations, where risk levels are calibrated to context-specific collision rate benchmarks, are declared Escalated Issues (EI). EIs are tracked for resolution (e.g., number of open EIs) within specified time targets (e.g., tracking closures and delays), with satisfactory program status predicated upon review and approval of each EI and conformance with resolution time targets. Maps to: architectural, behavioral, operational layers.

## (12) Field Safety:

*Risk Acceptance Criterion*: Continuous monitoring of potential sources of harm from ADS field operations is carried out to identify potentially new or not previously accepted sources of risk. Such issues are evaluated for immediate action — including complete or partial suspension of the Service — appropriate to the extent and severity of possible consequences. The live risk management function performed by the Field Safety program is complemented by regular evaluation of the Field Safety program health and maturity. Maps to: architectural, behavioral, operational layers.

*Discussion*

The previous section presented a total of twelve individual criteria. These criteria ground the structure (i.e., branches) of Waymo's Safety Case, where each criterion is complemented by a hierarchical decomposition of claims that argue the criterion reasonableness as well as the ability of a given methodology to credibly showcase their implementation [16]. As briefly noted in the Introduction, not every acceptance criterion is aimed at evaluating performance of the ADS. Process-based criteria are also accounted for, where rigor and completion status of specified activities inform readiness determination and eventual approval of a deployment.[10] Examples of those are found in the System Safety category, as well as in Verification and Validation, Risk Management, and Field Safety.

Not all criteria reach the same level of maturity at the same time, and the definition and implementation of each may follow a unique path of refinement. Furthermore, maturity and refinement of the data sources and data post-processing for the evaluation of each criterion also follow their unique development path (e.g., starting from manual triaging and analysis and then moving toward more robust automated features). It is thus important to ensure shared awareness of the status of each criterion and the data pipeline that feeds into it to appropriately inform clear governance practices (as discussed in more detail in the next section).

The actual computation and evaluation of the criteria is based on a number of data sources that combine testing data related to the particular configuration that is being considered for deployment, as well as deployment data related to previously approved configurations that remain behaviorally relevant and in scope (e.g., to understand trends, regressions, and improvements). Reliance on each data source type may change with time, also depending on the scope of changes associated with the ADS capabilities [17]. The usage of each data type, in turn, may also lead to different *degrees of confidence* in the predicted performance for the new release [17].

Taken as a whole, the above considerations imply that certain criteria may carry a higher weight than others. At Waymo, it is often the case that newer (or newly revised) criteria enter the Safety Framework (and the related documentation within the Safety Case) on an informative basis. Novel methodological developments for the evaluation of ADS safety follow what amounts to an internal "graduation process", which, when appropriate and based on the input from cross-functional stakeholders, may eventually lead to their inclusion into the formal readiness review process. As such, it is important to note that many additional indicators of readiness exist, beyond the formal criteria of the prior section. In fact, a number of other metrics provide additional color and information to guide the recommendation and the ultimate decision-making for approval.[11] Beyond the signal obtained from the informative methodologies described above, parallel reviews of operational readiness and incident management readiness ensure that the ADS service provider is able to support the scope and capabilities enabled by the latest software release (e.g., the expansion to freeway operations in addition to surface streets). Additional reviews may also be established ad-hoc, for example in situations in which a substantial platform modification (e.g., a new sensor suite, or a new model of the base vehicle) takes place, to complement the product evaluation workstreams anchored in the criteria of the prior section.

Finally, the readiness review process based on the presented criteria occurs at discrete points in time, grounding the approval of a specific configuration of the ADS for a specific deployment [21]. Such a stage is the natural culmination of a development cycle within the ADS iterative development model. When mapped to the Safety Determination Lifecycle of [16], this phase corresponds to the analysis of safety "as an acceptable prediction and/or observation". Prior to reaching such a stage, the implementation and checking of rigorous engineering practices guide what was termed "safety as an emergent development property" in [16]. A review of those practices, along with those that enable continuous confidence growth based on in-use monitoring data, are not in scope for this publication.

---

[10] This is *not* intended as a parallel to the notion that acceptance criteria can be both qualitative or quantitative (as defined in the SOTIF standard [20]), but rather as an orthogonal dimension. As such, one may have quantitative criteria to evaluate procedural rigor or *process* performance and completion and, similarly, qualitative criteria for *ADS* performance evaluation. It is important that such attributes remain distinct from each other, understanding the notion of procedural acceptance criteria as a complement to the (perhaps more intuitive) ADS performance-based ones that have populated recent literature.

[11] A notable example is the collection of metrics included in what Waymo terms the "hillclimbing report", provided as an informative complement at the time of a deployment readiness review. While there is no standardized definition of an hillclimbing metric, the notion is that of identifying frequently measurable performance indicators that provide, through comparison between each release, information that confirms that a given SW change/update leads to a directionally appropriate performance improvement and, as applicable, behaves as expected. In other words, these metrics allow setting directional goalposts that the software performance can be moved toward (a "hill to climb"). This iterative exploration is *not* intended to imply a formal optimization problem with the metric used as cost function.

## Decision-Making and Systematic Safety Governance

The readiness determination is, at its core, a risk assessment process. It is aimed at evaluating whether the deployment of a new release candidate, across the fleet and in specified locations, will pose a level of residual risk that remains in line with the pre-specified targets. What may often be termed "approval" is actually a much more nuanced decision than a binary pass/fail determination. In fact, the decision-making that takes place upon readiness determination answers both the question of whether the release is suitable for deployment (i.e., the "what") as well as the modality in which such deployment shall occur (i.e., the "how"). While we referred before to the deployment decision as the natural culmination of a development cycle, this stage should truly be considered an integral part of the risk mitigation tools available to the company management. A number of factors can be employed to ensure residual risk is appropriately controlled and managed during the approval stage, including;

- Appropriately determining which novel and/or revised attributes of an Operational Design Domain (ODD) should be considered in scope for the release. Controlling the ODD expansion and deployment plan is a clear approach to manage residual risk, especially in relation to newer capabilities (e.g., operations in denser fog). Purposeful scaling of the ODD goes hand in hand with the analysis of impact to residual risk stemming from such changes, where confidence in the ADS performance guides the decision to include or exclude certain ODD attributes from the scope definition attached to the release.
- Appropriately setting gradual rollout windows for the new release, with a phased deployment across the fleet and in each geography. This is done to better manage and control any unexpected behavior from the release, where the rollout is monitored and executed according to a pre-specified plan for each release. Additional monitoring capabilities during the initial rollout help reveal whether the ADS's performance aligns with the predictions on which the readiness review and approval were based [16].
- Appropriately setting the scale of the deployment size, as defined by the maximum mileage allowed for the fleet. The actual residual risk computation depends on both the probability of occurrence of harm and the consequences/severity of harm. Setting allowable mileage ranges across the fleet enables to bound exposure, which in turn controls residual risk levels.
- Appropriately distributing allowed mileage across multiple locations of deployment. This factor is grounded in the proper understanding and qualification of each operating environment. While the ADS has a unique Operational Design Domain (ODD) [1] in scope for the release, the instantiation of the ODD attributes in each location of deployment may vary (e.g., while rainfall within a certain downpour range may be in scope, the precipitation level experienced in City A vs. City B may be different, with the uncertainty in atmospheric conditions impacting residual risk differently across the two locations). One of the risk mitigation tools exercised upon approval is thus the selection of distribution of the mileage allowed in each geo. Such a decision is not undertaken to disfavor any one location of deployment; rather, it ensures that the technology rollout appropriately accounts for areas of maturity, areas of known limitations, and areas of uncertainty in a way that can be appropriately integrated within the existing geographic ecosystem as the ODD and system capabilities get purposefully expanded.

As mentioned before, a number of additional metrics and parallel review processes (including health monitoring capabilities for real-time incident escalation and management) ensure that the approval decision is undertaken holistically. No single methodology or criterion is sufficient to determine absence of unreasonable risk on its own, just like the determination of an unreasonable level of risk is seldom associated with a single failing indicator.[12] Rather, the combination of the twelve criteria here presented enables a qualitative aggregation of residual risk, informed by the evaluation of each criterion against individual methodologies' targets. As presented in [21], continuous monitoring of on-road performance, with the ability to intervene on operations whenever deemed necessary, provides added confidence in the ability to both evaluate and manage the event-level performance of the system, in addition to the qualitative aggregation done during the approval process.

Targets associated with each methodology and criterion are pre-specified (and/or reconfirmed) ahead of each evaluation cycle. Changes in methodological processes and/or modifications of the acceptance criteria (e.g., to enable analysis of a broader scope of a given release) require revisions or reconfirmation of respective targets, and would follow a gradual implementation process as previously discussed. Criteria targets are based on the extensive subject matter expertise of those involved in the assessment. Targets' setting also uses a dedicated process for definition and approval, so that a clear governance structure plays a fundamental role well before the readiness determination stage. Roles and responsibilities across various tiers of stakeholders along the processes that lead to readiness determination are schematically represented in Figure 1.

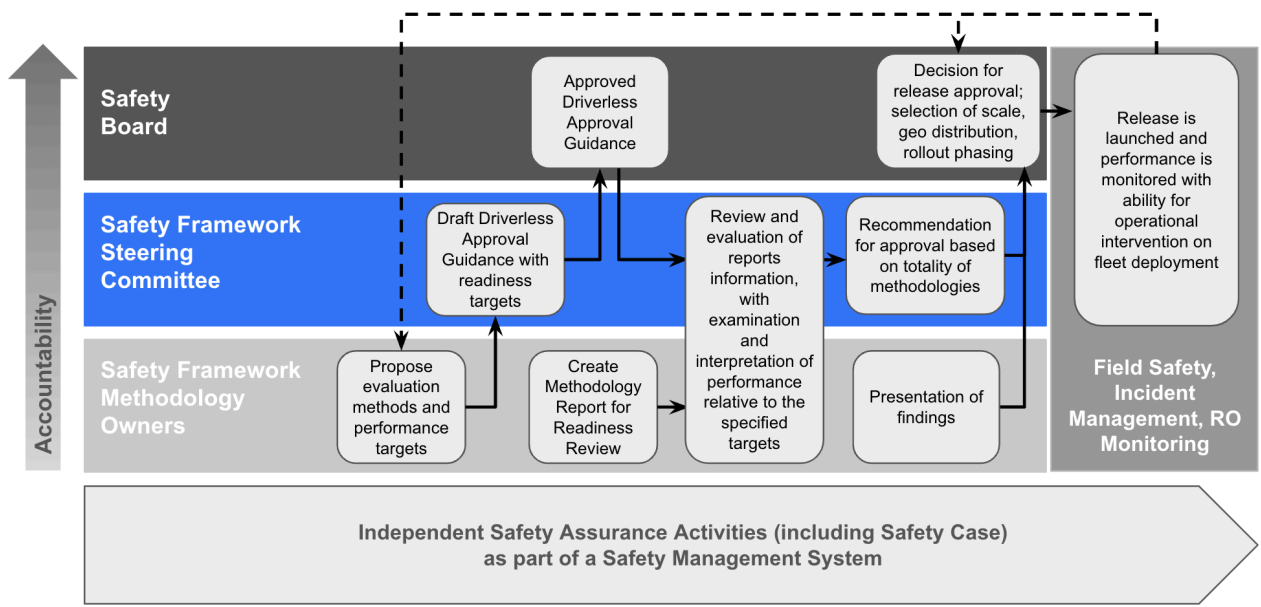

Figure 1. Schematic representation of roles and responsibilities toward readiness determination for various tiers of internal stakeholders.

Figure 1 identifies three tiers of increasing accountability for approval:

- Safety Framework Methodology Owners: each of the methodologies (and associated criteria) that populate Table 1 is assigned a formal owner, who oversees the implementation of the methodology pipeline and retains the technical expertise necessary to ensure the appropriate level of rigor is maintained across the organization and execution of the approach. Their intimate familiarity with the methodology structure makes this category of stakeholders ideally suited to elevate recommendations on performance targets as well as establish proposals on any needed modifications to the product evaluation processes. During an active readiness review phase, methodology owners are also responsible for the drafting of a methodology report catered to the particular release being evaluated, as well as its subsequent presentation to the other categories of stakeholders.

---

[12] This does not mean that a single event may not be capable of blocking an entire deployment or lead to a restriction of an active deployment - both of which have in fact happened in the past. Rather, it implies that no criterion dominates the others and that no formal gating prioritization currently exists across the criteria of Table 1.

- Safety Framework Steering Committee: a cross-functional group of senior leaders (across both technical and management functions) is responsible for reviewing the recommendations and reports from each methodology owner to formalize: (i) guidance on readiness targets for each planned release; and (ii) upon an actual release review stage, draft the recommendation for approval following the holistic assessment of the input provided by each methodology owner.
- Safety Board: the final tier retains accountability for the overall readiness determination process, and is responsible for the approval of both the targets' guidance (with or without modifications compared to the recommendation from the steering committee - item (i) above), as well as the actual release approval. As discussed previously, approval also entails the fine tuning of a number of factors that enable managing and controlling residual risk across the fleet, including allowed deployment scale, distribution of the mileage across multiple locations of deployment, and plans for phased rollout.

Following the decision for approval, the operational launch of the fleet is closely monitored and followed by a number of internal teams, ready to act quickly to address any safety issue that may arise post-deployment. Constant monitoring and Field Safety review of any novel concern reveal whether the ADS's performance aligns with the targets on which the readiness review and approval were based and can inform continuous improvement for the process (see the dashed feedback lines at the top of Figure 1). A final category of internal stakeholders exists, albeit implicitly visualized in Figure 1. Those are the individuals responsible for independent assurance activities (represented to the bottom of Figure 1), including those pertaining to the documentation of the readiness process, its justification and pressure testing within the Safety Case.[13]

Throughout the readiness determination process and its supporting activities, open communication across the various stakeholders is of utmost importance. Beyond the in-depth review and discussion[14] of methodology reports elevated during the readiness review, direct access to technical experts for questions and comments resolution is available to all senior company leaders. This process does not stop upon readiness determination: following approval, Field Safety processes and in-use monitoring feed information back to continuously oversee system performance, confirm pre-deployment estimates, and inform improvements[15] toward subsequent development. Independent activities of safety assurance, including the Safety Case itself, aided by the cross-functional nature of all the tiers (involving engineering, product, operations, safety, and legal divisions working together) support the entire governance process by ensuring appropriate documentation of methodology processes and assumptions, supporting the validation of the stated acceptance criteria and associated targets, and confirming the appropriate implementation of the intended processes.

## Limitations

The criteria presented in this paper are not intended to be sufficient for the determination of absence of unreasonable risk. The implementation of each criterion, as well as their holistic assessment, requires a robust set of processes, resources, and tooling whose presentation is outside the scope of this paper but that remain an integral part of the safety determination. This recommendation is based on a wealth of lessons from the implementation of Waymo's RO readiness review process for the past several years, where the evaluation approaches at the basis of each methodology have undergone a number of revision cycles. At the basis of appropriate governance for readiness determination stands a responsible approach to safety management, so that care should be exercised by anyone attempting to replicate the criteria reviewed in this paper to ensure the appropriate complementary processes are also in place.

---

[13] As noted in [16], independent development of the safety case documentation separate from those who own the respective methodologies and processes aids the identification of epistemic defeaters and reduces the potential for confirmation bias. We have optimized for accuracy, efficiency, and a balance in expertise between external best practices and internal process, developing a safety case through a team led by certified safety experts, who were independent from those who developed and/or executed the Safety Framework. Review and revisions of the safety case content are carried out by subject matter experts internal to the company (including the owners of each evaluative methodology part of the Safety Framework) to help ensure that the safety case developers have correctly represented a methodology description and its coverage, limitations, etc. Finally, a third internal independent reviewer is used as an additional layer of review to provide an independent assessment against bias in the development of the safety case.

[14] This includes a day-of official presentation from each methodology owner and key subject matter experts.

[15] This is not intended to exclusively refer to unplanned changes/improvements, but also to the collection of information that supports planned improvements for future software releases (e.g., to enable expanded capabilities for the system, which are typically mapped to future releases months in advance).

As stated before, the presentation of the criteria for readiness determination is also not an indication of a finalized status of a company's approach, where methodologies, criteria, related indicators, and product capabilities remain in continuous evolution. The intent of this publication is rather to continue to further the ongoing conversation on ADS safety with a more concrete proposal for a minimum set of criteria that can ground the determination of Absence of Unreasonable Risk. This type of proposals aid explainability and understanding of the processes that lead a company to deploy this technology on public roads with enough confidence on their ability to be able to predict the residual risk associated with a new configuration while at the same time bounding uncertainty through the ability to rapidly respond to safety issues that may arise.

## Conclusions

This paper is intended as a companion to many preceding publications, both from Waymo (e.g., the presentation of readiness methodologies [21], the safety case approach white paper [16]) and from other stakeholders, be they ADS companies (e.g., Aurora, Gatik), industry consortia (e.g., AVSC [26]), standard development organizations (e.g., UL [9], SAE [27]), or regulatory agencies (e.g., [10], [11]) that have collectively shaped the conversation on safety determination. A review of those publications may aid the reader in understanding a more cohesive picture of the proposed approach to safety and the role and contribution of this paper toward practical minimum criteria for the determination of absence of unreasonable risk. Other publications may be considered as individual pieces of evidence within the broader safety case, and speak to or fulfill different functions. For example: papers connected to the retrospective analysis of performance [13] [14] [28] [29] speak to the "continuous confidence growth" portion of the Safety Determination Lifecycle [16]; methodological papers on collision avoidance [23] [24] substantiate claims connected to one of the scenario-based testing methodologies of Table 1; examples of benchmarking approaches [22] [25] aid the understanding of targets' setting for one of the simulated deployment methodologies of Table 1. While, collectively, each of those publications informs important aspects of a safety determination, a cohesive framing of minimum performance indicators toward approval of a given ADS configuration has so far not been agreed upon by the many categories of stakeholders touched by ADS technology. As such, in this paper we proposed a set of twelve criteria for the operationalization of the determination of Absence of Unreasonable Risk that we hope can inform the continued discussion on this important topic within standardization and regulatory contexts.

The methodological criteria presented were mapped and organized within Table 1 according to traditional system engineering areas of system development and testing and mapped to the activities presented in Waymo's first publication of readiness methodologies [21]. Those criteria are connected with: 1) System Safety; 2) Cybersecurity; 3) Verification and Validation; 4) Collision Avoidance Testing; 5) Predicted Collision Risk;  6) Impeded Progress; 7) Rules of the Road Compliance; 8) Vulnerable Road Users Interactions;  9) High-Severity Assessment; 10) Conservative Severity Estimate; 11) Risk Management; and 12) Field Safety. While the criteria have been informed by Waymo's extensive experience in this domain, they are here formulated at a level of abstraction suitable for an industry-wide implementation and evaluation regardless of the specific technological solution and/or architecture tackled by individual developers. They are, in this sense, technology- and implementation-agnostic, and can inform and further the conversation between industry and regulatory stakeholders toward the creation of minimum safety standards. While foundational standardization work that grounds definition and understanding of ADS technology is well under-way, additional work is needed on a more precise operationalization of the definition of absence of unreasonable risk. This paper can thus serve as an important contribution toward the creation of such future standards, also furthering the conversation with policy-makers and regulators on how to ensure overall public safety while, at the same time, enabling the transformative promise of ADS technology.

## Acknowledgments

Many others across Waymo have contributed to the creation and refinement of Waymo's Safety Framework Methodologies and Safety Case, as well as to the review and conceptualization of the content in this paper.

## References

[1]. International Organization for Standardization (ISO) ISO/SAE PAS 22736:2021 Taxonomy and definitions for terms related to driving automation systems for on-road motor vehicles

[2]. Blumenthal, M.S., Fraade-Blanar, L., Best, R. and Irwin, J.L.,(2020). Safe enough: Approaches to Assessing Acceptable Safety for Automated Vehicles. Rand Corporation.

[3]. Koopman, P. (2022). How safe is safe enough? Measuring and predicting autonomous vehicle safety.

[4]. Salem, N.F., Kirschbaum, T., Nolte, M., Lalitsch-Schneider, C., Graubohm, R., Reich, J. and Maurer, M., 2024. Risk Management Core–Towards an Explicit Representation of Risk in Automated Driving. *IEEE Access*.

[5]. Favarò, F. (2021). Exploring the Relationship Between" Positive Risk Balance" and" Absence of Unreasonable Risk". *arXiv preprint arXiv:2110.10566*

[6]. Sandblom, F., De Campos, R., Warg, F., Beckman, F., Choosing Risk Acceptance Criteria for Safety Automated Driving, 2024. In: European Dependable Computing Conference (EDCC) 2024.

[7]. Automated Vehicle Safety Consortium. 2024. AVSC Best Practice for Core Automated Vehicle Safety Information - AVSC-D-02-2024. SAE Industry Technologies Consortia.

[8]. International Organization for Standardization (ISO) Road vehicles — Safety for automated driving systems — Design, verification and validation ISO/AWI TS 5083. In drafting.

[9]. Underwriters Laboratories (UL) - UL 4600:2023 Standard for Safety Evaluation of Autonomous Products - Third Edition. (2023)

[10]. Regulation (EU) 2022/1426 - https://eur-lex.europa.eu/legal-content/EN/TXT/HTML/?uri=CELEX:32022R1426

[11]. GRVA 18-50. Guidelines and recommendations for ADS safety requirements, assessments and test methods to inform regulatory development. Available at: https://unece.org/sites/default/files/2024-01/GRVA-18-50e_1.pdf

[12]. Koopman, P. and Widen, W.H., 2023. Breaking the Tyranny of Net Risk Metrics for Automated Vehicle Safety. *Available at SSRN 4634179*.

[13]. Kusano, K.D., Scanlon, J.M., Chen, Y.H., McMurry, T.L., Chen, R., Gode, T. and Victor, T., 2023. Comparison of Waymo Rider-Only Crash Data to Human Benchmarks at 7.1 Million Miles. *arXiv preprint arXiv:2312.12675*.

[14]. Di Lillo, L., Gode, T., Zhou, X., Atzei, M., Chen, R. and Victor, T., 2023. Comparative safety performance of autonomous-and human drivers: A real-world case study of the Waymo One service. *arXiv preprint arXiv:2309.01206*.

[15]. Chen, J.J. and Shladover, S.E., 2024. Initial Indications of Safety of Driverless Automated Driving Systems. *arXiv preprint arXiv:2403.14648*.

[16]. Favaro, F., Fraade-Blanar, L., Schnelle, S., Victor, T., Peña, M., Engstrom, J., Scanlon, J., Kusano, K. and Smith, D., 2023. Building a Credible Case for Safety: Waymo's Approach for the Determination of Absence of Unreasonable Risk. *arXiv preprint arXiv:2306.01917*.

[17]. Favaro, F.M., Victor, T., Hohnhold, H. and Schnelle, S., 2023. Interpreting Safety Outcomes: Waymo's Performance Evaluation in the Context of a Broader Determination of Safety Readiness. 9[th] International Symposium on Transportation Data & Modelling (ISTDM2023). Ispra, June 19-22 2023, available at pg. 364: https://publications.jrc.ec.europa.eu/repository/handle/JRC134973

[18]. International Organization for Standardization (ISO). (2018). Road Vehicles - functional safety ISO 26262:2018.

[19]. United States Code, 2018 Edition, Supplement 4, Title 49 - TRANSPORTATION. SUBTITLE VI - MOTOR VEHICLE AND DRIVER PROGRAMS PART A - GENERAL CHAPTER 301 - MOTOR VEHICLE SAFETY. Sec. 30102 - Definitions. Available at: https://www.govinfo.gov/content/pkg/USCODE-2022-title49/pdf/USCODE-2022-title49-subtitleVI-partA-chap301.pdf

[20]. International Organization for Standardization (ISO). (2022). Road Vehicles - safety of the intended functionality ISO 21448:2022.

[21]. Webb, N., Smith, D., Ludwick, C., Victor, T.W., Hommes, Q., Favarò F., Ivanov, G., and Daniel, T. (2020). Waymo's Safety Methodologies and Safety Readiness Determinations. *arXiv preprint  arXiv:2011.00054*

[22]. Scanlon, J.M., Kusano, K.D., Fraade-Blanar, L.A., McMurry, T.L., Chen, Y.H. and Victor, T., 2023. Benchmarks for Retrospective Automated Driving System Crash Rate Analysis Using Police-Reported Crash Data. *arXiv preprint arXiv:2312.13228*.

[23]. Kusano, K.D., Beatty, K., Schnelle, S., Favaro, F., Crary, C. and Victor, T., 2022. Collision avoidance testing of the waymo automated driving system. *arXiv preprint arXiv:2212.08148*.

[24]. Engström, J., Liu, S.Y., DinparastDjadid, A. and Simoiu, C., 2024. Modeling road user response timing in naturalistic traffic conflicts: a surprise-based framework. *Accident Analysis & Prevention*, *198*, p.107460.

[25]. Scanlon, J.M., Teoh, E.R., Kidd, D.G., Kusano, K.D., Bärgman, J., Chi-Johnston, G., Di Lillo, L., Favarò, F.M., Flannagan, C.A., Liers, H. and Lin, B., 2024. RAVE checklist: Recommendations for overcoming challenges in retrospective safety studies of automated driving systems. Traffic Injury Prevention, pp.1-14.

[26]. Automated Vehicle Safety Consortium (AVSC). 2024. Best Practice for Core Automated Vehicle Safety Information. SAE Industry Technologies Consortia.

[27]. SAE International. 2025. Driving Assessment (DA) Metrics for Automated Driving Systems. SAE J3237 *(in publication)*.

[28]. Victor, T., Kusano, K., Gode, T., Chen, R., Schwall, M., 2023. Safety performance of the waymo rider-only automated driving system at one million miles. *Available at waymo.com/safety/research*

[29]. Di Lillo, L., Gode, T., Zhou, X., Scanlon, J. M., Chen, R., & Victor, T. (2024). Do Autonomous Vehicles Outperform Latest-Generation Human-Driven Vehicles? A Comparison to Waymo's Auto Liability Insurance Claims at 25.3M Miles.

[30]. Chen, Y., Scanlon, J. M., Kusano, K. D., McMurry, T., Victor, T. (2024). Dynamic Benchmarks: Spatial and Temporal Alignment for ADS Performance Evaluation. arXiv preprint arXiv:2410.08903. https://arxiv.org/pdf/2410.08903